# Introduction of accelerated BOIN design and facilitation of its application

Masahiro Kojima, Ph.D.[1]

Wu Wende, MSc.[2]

Henry Zhao, Ph.D.[2]

[1]Biometrics Department, R&D Division, Kyowa Kirin Co., Ltd., Tokyo, Japan.

[2]Biometrics Department, Development Division, Kyowa Kirin, Inc., Princeton, NJ, USA

**Running title**: accelerated BOIN design

**Financial support**: None

**Corresponding author**

Name: Masahiro Kojima, Ph.D.

Address: Biometrics Department, R&D Division, Kyowa Kirin Co., Ltd.

Otemachi Financial City Grand Cube, 1-9-2 Otemachi, Chiyoda-ku, Tokyo, 100-004, Japan.

Tel: +81-3-5205-7200

FAX: +81- 3-5205-7182

Email: masahiro.kojima.tk@kyowakirin.com

**A conflict of interest disclosure statement**: None

**Word count**: 3119

**The number of figures**: 3

**The number of tables**: 2

Funding: Not applicable.

Authors' contributions: M. Kojima: Conception and design; development of methodology; acquisition of data (provided animals, acquired and managed patients, provided facilities, etc.): analysis and interpretation of data (e.g., statistical analysis, biostatistics, computational analysis); writing, review, and revision of the manuscript; administrative, technical, and material support (i.e., reporting and organizing data, constructing databases); and study supervision. W. Wende: review. Z. Henry: Conception and design; analysis, writing, review.

Acknowledgements: MK thanks Professor Hisashi Noma for his encouragement and helpful suggestions.

Patients are directly involved in this study: Not applicable.

**150-word statement of translational relevance:**

**Key objective:** Can a clinical trial designer suggest a BOIN design to clinicians who only have experience with the 3+3 design that would complete the study as quickly as the 3+3 design while increasing the accuracy of correct maximum tolerated dose (MTD) selection?

**Knowledge generated:** We propose an accelerated BOIN design to be completed in a phase I trial as quickly as the 3+3 design. We also introduce how we were the first to apply the BOIN design where our company applied the 3+3 design for most of the clinical oncology dose-escalation trials.

**Relevance:** The information we present is useful for those who want to finish an accelerated BOIN design as quickly as a 3+3 design or who are having trouble applying a BOIN design.

**Abstract**

**Purpose:** During discussions at the Data Science Roundtable meeting in Japan, there were instances where the adoption of the BOIN design was declined, attributed to the extension of study duration and increased sample size in comparison to the 3+3 design. We introduce an accelerated BOIN design aimed at completing a clinical phase I trial at a pace comparable to the 3+3 design. Additionally, we introduce how we could have applied the BOIN design within our company, which predominantly utilized the 3+3 design for most of its clinical oncology dose escalation trials.

**Methods**: The accelerated BOIN design is adaptable by using efficiently designated stopping criterion for the existing BOIN framework. Our approach is to terminate the dose escalation study if the number of evaluable patients treated at the current dose reaches 6 and the decision is to stay at the current dose for the next cohort of patients. In addition, for lower dosage levels, considering a cohort size smaller than 3 may be feasible when there are no safety concerns from non-clinical studies. We demonstrate the accelerated BOIN design using a case study and subsequently evaluate the performance of our proposed design through a simulation study.

**Results**: In the simulation study, the average difference in the percentage of correct MTD selection between the accelerated BOIN design and the standard BOIN design was -2.43%, the average study duration and the average sample size of the accelerated BOIN design was reduced by 14.8 months and 9.22 months, respectively, compared with the standard BOIN design.

**Conclusion**: We conclude that our proposed accelerated BOIN design not only provides superior operating characteristics but also enables the study to be completed as fast as the 3+3 design.

**Introduction**

Dose escalation studies in oncology play a pivotal role in determining the safe therapeutic doses for patients. In recent years, the emergence of model-assisted designs has gained significant traction because of their simplicity like the algorithm-based designs and high accuracy like the model-based designs[1]. Particularly, the Bayesian Optimal Interval (BOIN) design stands out for easy implementation and has been recognized extensively in clinical research[2]. A recent search on Clinicaltrials.gov on August 14, 2023, with the keywords "Bayesian optimal interval or BOIN" yielded 133 trials, underscoring the fast adoption of the BOIN design in the clinical trials. Further endorsing its credibility, the BOIN design has received a fit-for-purpose designation from the FDA[3]. There have also been numerous derivative designs stemming from the standard BOIN model[4].

The Data Science Roundtable (DSRT) meeting, convened in Japan, serves as a platform for deliberation on statistical challenges associated with drug development. The DSRT meeting gathering brings together stakeholders from pharmaceutical companies, the regulatory authority (PMDA: Pharmaceuticals and Medical Devices Agency), and academia. The 7th DSRT meeting covered three themes[5], one of which was associated with novel phase I clinical trials in oncology. Eight topics formed the backbone of discussions on novel phase

I clinical trials, with participants delving deep into each subject matter. For the reader, the English translations of these topics from Japanese are provided in the supplemental material.

Discussions at the DSRT meeting highlighted practical challenges associated with the application of the BOIN design. Distinct from the 3+3 design, the BOIN design needs to continue the dose-escalation study until a predetermined sample size is achieved, typically between 20 to 30 participants. The sample size of the BOIN design often exceeds that of the more commonly used 3+3 design. Consequently, there is hesitancy in adopting the BOIN design, given concerns about potential extensions in study duration and cost increases when compared to the 3+3 design. Another challenge is the accuracy metric used by the BOIN design, which is the percentage of correct MTD selection. Clinicians accustomed to the 3+3 design are often unfamiliar with this metric. For instance, the assertion that the BOIN design augments the percentage of correct MTD selection by a significant 20% compared to the 3+3 design may not convey its intended significance due to this unfamiliarity. While large pharmaceutical corporations or major hospitals, experienced in model-based designs such as the CRM, might transition to the BOIN design with relative ease, the same cannot be said for smaller entities. Small to medium-sized companies or hospitals, having predominantly relied on the 3+3 design, may find migrating to the BOIN framework daunting. Their established comfort and operational fluency with the 3+3 design make the switch to the BOIN design less enticing. This transition is further complicated by a discernible knowledge gap, wherein these organizations struggle to articulate the advantages of the BOIN design and strategize its effective integration.

In this paper, we introduce the accelerated BOIN design, which allows to complete the dose escalation study as quickly as the 3+3 design while significantly increasing the accuracy of MTD selection. We work for a small sized pharmaceutical company. Our company has only used 3+3 designs. Recently, we have been approved to apply the BOIN design in our company as the first project team. We will introduce this experience to show how we have applied the BOIN design. If a reader is familiar with BOIN design well, one may feel that the design presented here is an obvious one, but we found at the DSRT meeting that it is surprisingly not well understood, which hinders the spread of the BOIN design. Hence, we prepare this paper aiming at demonstrating and facilitating the application of the accelerated BOIN Design.

## Methods

First, we review the BOIN design. The BOIN design gives the optimal dose adjustment rule by assuming three assumptions: true DLT probability is less than the target toxicity level (TTL), true DLT probability is equal to TTL, and true DLT probability is greater than TTL, and by considering the problem of minimizing the sum of three probabilities of each assumption being judged incorrect. For example, for TTL=0.3, the boundaries for dose adjustment are (0.236, 0.358). If the observed DLT rate is under 0.236 then the dose for the next cohort escalates, if it is between 0.236 and 0.358 then the dose for next cohort retains, if it is over 0.358 then the dose for next cohort de-escalates. Based on the boundaries, we can pre-tabulate a table with the number of DLTs for dose escalation/de-escalation decision. There is a stopping option in BOIN design to stop the dose-escalation trial when enough patients have been administered at a dose level. The standard cohort size is 3, but unlike the algorithm-based 3+3 design, the cohort size can be smaller or larger than 3 in the BOIN design.

In this paper, we introduce a design that takes advantage of the stopping option of the BOIN design and the feature to adjust cohort size to complete the trials as quickly as the 3+3 design. Because the maximum number of patients treated at the current dose in the 3+3 design is six, we propose that the stopping rule for Accelerated BOIN design to terminate the dose-escalation trial is when the number of evaluable patients treated at the current dose reaches 6 and the decision is to stay at the current dose. To our knowledge, there are no studies that have made simulation comparisons with the 3+3 design under our proposed stopping rule, nor are there any studies that have applied our proposed approach. Regarding the cohort size, it should be smaller than 3 if the results of non-clinical and previous clinical trials suggest that the probability of DLT is very small at low doses. In a phase 1/2 trial, Gainor et al. (2021)[6] applied the cohort size of 1 for the lowest dose in BOIN design.

**Case study**

We demonstrate through a case study referring to VIOLA trial (a prospective, open-label, dose-finding, multicenter, phase I trial to determine the MTD of lenalidomide (LEN) in combination with azacytidine (AZA))[8]. The doses administered in the VIOLA study were 5 mg LEN with AZA (starting dose), 10 mg LEN with AZA, 15 mg LEN with AZA, and 25 mg LEN with AZA. DLT occurred at the 25 mg LEN with AZA, one in the second cohort and one in the third cohort. We consider 4 dose level in this case study. The DLT evaluation period is 84 days. The accrual time is 30 days per patient. The standard cohort size is three.

The maximum sample size of the BOIN design is 21. The TTL is 20% and the boundaries of dose escalation/de-escalation is (0.157, 0.238).

The accelerated BOIN design can complete the trial early when the number of evaluable patients treated at the current dose reaches 6 and the decision is to stay at the current dose for the next cohort. We assume that the DLT probability at dose 1 and dose 2 is very low based on the results of non-clinical studies.

We illustrate the process of dose escalation with accelerated BOIN design, standard BOIN design, and the 3+3 design in Figure 1. For cohort 1 and cohort 2, the accelerated BOIN design has a smaller cohort size than the other two designs and has been able to shrink the trial by 4 months to reach cohort 3. For the accelerated BOIN design, Cohort 6 would stay at dose 4 after 6 patients from cohort 4 and 5 were treated at that dose level, per proposed stopping rule the trial can be stopped early as in the 3+3 design. However, the accelerated BOIN design enables the study to be completed 4 months earlier than the 3+3 design. Whereas the standard BOIN design continued dosing until the sample size of 21 was reached, finishing the study 15 months later than the accelerated BOIN design.

We conducted a numerical simulation study to show that the accelerated BOIN design performed better than the 3+3 design and comparably to the BOIN design.

**Numerical simulation study.** We conducted a simulation study to evaluate the performance and operating characteristics of the accelerated BOIN design compared to the 3+3 design and the standard BOIN design. We considered a dose finding trial with 5 dose levels, a sample size of 24 patients, and two target toxicity levels of 25% and 30%. For each TTL, six toxicity scenarios were investigated with 10,000 simulated trials under each scenario. The true DLT probabilities are shown in Supplemental Material. The safety stopping rule for overdose control proposed by Liu and Yuan (2015) was applied. For the accelerated BOIN design, the initial cohort size of 1 for dose level 1 and dose level 2 was used for all scenarios. We assumed that the accrual time for each patient was 30 days and the DLT evaluation period was 84 days.

We evaluated the following 4 criteria via simulations.

1. Percentage of correct MTD selection

2. Study duration

3. Total sample size

4. Percentage of stopping

**Promotion of the Application of the BOIN Design**

We have been interested in the flexibility and accuracy of the correct MTD selection provided by the BOIN design, attributes that are lacking in the 3+3 design. Consequently, we explain how we could proceed to apply the BOIN design in an environment where only the 3+3 design was used previously.

First, consultation with the clinician was essential to facilitate the implementation of the BOIN design. At the beginning, the clinician was apprehensive, predominantly due to the unfamiliarity with the dose adjustment methodology, which differs from the 3+3 design they were accustomed to. Moreover, they concerned about the potential operational challenges associated with the BOIN design, given their extensive experiences with the 3+3 design.

To address the concerns, we created dose-adjustment transition diagrams (such as Figure 1) for the various patterns to provide a clear representation of how doses would be changed under various DLT occurrence scenarios. Stakeholders in drug development, including clinicians and clinical pharmacology, felt that with the 3+3 design, they only needed to be concerned with the number of DLT occurrences related to the most recent dose.

In contrast, with the BOIN design the dose adjustment method of repeatedly increasing and decreasing doses was complicated. For example, it was difficult for non-statisticians to understand how dose adjustment was made at DLT evaluation of dose level 1 after escalating to dose level 2 and then de-escalating back to dose level 1, why the dose adjustment was based on all DLT assessment results of dose level 1, including the result from previous assessment at dose level 1. The first explanation of the BOIN design was met with resistance due to their familiarity with the 3+3 design. However, after multiple attempts, the resistance faded and they became interested in the BOIN design. Also, it is important to keep the perception of the BOIN design in line within the statistical analysis personnel in advance. As a second opinion, the clinician also checked with other biostatisticians who were not on the study to see if the BOIN design could be operated without any problems. After several explanations, we were able to get approval from clinician for applying the BOIN design. We also explained the BOIN design to the clinical pharmacology staff, medical writers and data management personnel to make sure they understood the operation of the BOIN design. To start a clinical trial, it is usually necessary to get approval from someone who is not in charge of conduct of the actual trial. We were allowed to participate in meetings with the approvers and explained the BOIN design to them. The clinician was also present and provided support as needed. We felt that it would have gone more smoothly if we had obtained the consent from the Head of the clinical sciences for the application of the BOIN design in advance.

One of the questions raised in the previous explanations was the increase in the number of patients treated and the longer study duration compared to the 3+3 design. Therefore, we proposed the accelerated BOIN design and told them that the trial can be completed earlier than the 3+3 design if no DLT is observed, and even if DLT is observed, the study period will be slightly longer than the 3+3 design, but the MTD will be estimated more accurately. We also firmly informed them of the advantages of enhancing the percentage of correct MTD selection, and the necessity and importance of correct MTD selection, and that it has a significant bearing on the probability of success of phase 2 and phase 3 from the paper by Conaway et al. (2018)[7]. Thus, Personnel involved in development, such as approvers, clinicians, medical writers, clinical pharmacology, and data management understood the importance of MTD estimation.

## Results

All simulation results are shown in Table 1 and Table 2.

[Percentage of correct MTD selection]

The visual results of percentage of correct MTD selection are shown in Figure 2 and Figure 3. For the target toxicity level of 30%, the average difference between the accelerated BOIN and standard BOIN (accelerated BOIN – standard BOIN) is -2.43% (minimum of -8.49% and maximum of 3.34%). The average difference between the accelerated BOIN and 3+3 (accelerated BOIN – 3+3) is 17.88% (minimum of 1.47% and maximum of 39.13%). For the target toxicity level of 25%, the average difference (accelerated BOIN – standard BOIN) is -2.58% (minimum of -6.17% and maximum of 1.26%). The average difference (accelerated BOIN – 3+3) is 15.64% (minimum of -0.84% and maximum difference of 43.30%).

[Average study duration]

The visual results of average study duration are shown in Figure 2 and Figure 3. For the target toxicity level of 30%, the average difference (accelerated BOIN – standard BOIN) is -14.08 months (minimum of -15.39 months and maximum of -11.58 months). The average difference (accelerated BOIN – 3+3) is 3.33 months (minimum of -4.03 months and maximum of 10.74 months). For the target toxicity level of 25%, the average difference (accelerated BOIN – standard BOIN) is -13.97 months (minimum of -15.79 months and maximum of -11.81 months). The average difference (accelerated BOIN – 3+3) is 2.12 months (minimum of -4.12 months and maximum of 8.90 months).

[Average total sample size]

The visual results of the average total sample size are shown in Figure 2 and Figure 3. For the target toxicity level of 30%, the average difference (accelerated BOIN – standard BOIN) is -9.22 (minimum of -9.80 and maximum of -7.92). The average difference (accelerated BOIN – 3+3) is -0.21 (minimum of -4.02 and maximum of 3.62). For the target toxicity level of 25%, the average difference (accelerated BOIN – standard BOIN) is -9.16 (minimum of -10.10 and maximum of -8.04). The average difference (accelerated BOIN – 3+3) is -0.83 (minimum of -4.06 and maximum of 2.67).

[Percentage of early stopping]

For both target toxicity levels of 30% and 25%, the percentage of early stopping of accelerated BOIN ranged from about 80% to 99%. The percentage of early stopping of standard BOIN ranged from about 0% to 13%. The percentage of early stopping of 3+3 ranged about 0% to 52%.

## Discussion

We introduced the accelerated BOIN design to complete dose-escalation trials as quickly as the 3+3 design. The accelerated BOIN design was set up to reduce the cohort size by better configuring the stopping option provided in the BOIN design. The DSRT meeting revealed that the BOIN design is frowned upon because of the relatively larger sample size and longer study duration compared with the 3+3 design. Therefore, we introduced how we could have applied the BOIN design within our company, which predominantly utilized the 3+3 design for most of its clinical oncology dose escalation trials.

In the case study, using DLT occurrences scenario in a real dose-escalation trial, our proposed design had the smallest sample size and shortest study duration. The reduced sample size may allow the addition of backfill cohorts for intermediate/high doses or to move to an extension part sooner.

The simulation study demonstrated that there was almost no difference in the percentages of correct MTD selection between our proposed design and the BOIN design for all scenarios, the accelerated BOIN design is comparable to the standard BOIN design. On the other hand, compared with the 3 + 3 design, our proposed design has sufficiently higher percentages of MTD selection for all scenarios except Scenario 1. We showed that our proposed design can lead to a significant reduction in both study duration and sample size compared with the BOIN design. When compared to the 3+3 design, our proposed design has smaller sample size and shorter study duration when the candidate MTD is on the high-dose side. Because the usual dose-escalation study sets the MTD at medium or higher doses, it is expected that the study would be completed sooner than the 3+3 design if the true MTD is as assumed. The percentage of early completion of the trial was very high due to the efficient setting of the stopping option in our proposed design. Although we completed the trial early, the accuracy of correct MTD selection does not change compared to the standard BOIN design.

The PMDA has an investigational new drug application checklist for the anti-cancer field, if a design other than the 3+3 design is selected for a dose-escalation trial, the following items must be verified in advance.

1. Definition of MTD, conditions for completion of tolerability evaluation (maximum number of patients expected), etc.

2. The results of a simulation-based study of which dose would be selected as the MTD if the probability of true DLT at the lowest dose level slightly exceeded the probability defined as hypertoxicity (the percentage of each dose selected), the average number of subjects at each dose, and the probability of study discontinuation.

We had an experience that the values recommended in the original paper of the BOIN design for trial discontinuation did not meet PMDA requirements. Because our proposed design would allow the trial to be terminated with high percentage, it may not be necessary to change the values recommended in the BOIN design paper.

We prepared a new simulation application with additional options for our proposed design in R shiny (https://masa-koji.shinyapps.io/accelerated_BOIN/). Because the BOIN design in the trialdesign.org[9] does not apply dose titration at dose level 1 only and does not show a study duration, a new simulation application was developed.

We have confirmed that the performance of our proposed design is superior. The accuracy of our proposed design is almost identical to the standard BOIN design. Additionally, our proposed design reduces the total sample size and study duration compared to the standard BOIN design. Therefore, we recommend the use of our design. If you consider applying the BOIN design, please refer to how we have applied it in the past.

Table 1 Simulation results for target toxicity level of 30%

| Design | %MTD | | | | | Ave SS | Ave SD | %ES |
|---|---|---|---|---|---|---|---|---|
| | 1 | 2 | 3 | 4 | 5 | | | |
| Scenario 1 | | | | | | | | |
| ABOIN | 0.01 | 0.01 | 0.12 | 0.49 | **99.37** | 14.28 | 30.35 | 99.94 |
| BOIN | 0.00 | 0.00 | 0.00 | 0.05 | **99.95** | 24.00 | 45.40 | 0.00 |
| 3+3 | 0.00 | 0.10 | 0.80 | 1.20 | **97.90** | 18.30 | 34.38 | 0.00 |
| Scenario 2 | | | | | | | | |
| ABOIN | 1.63 | 3.01 | 12.45 | 32.53 | **50.33** | 16.07 | 33.81 | 88.40 |
| BOIN | 0.32 | 1.66 | 8.89 | 36.57 | **52.53** | 23.99 | 45.39 | 0.03 |
| 3+3 | 6.10 | 9.00 | 23.80 | 37.30 | **20.20** | 18.70 | 35.12 | 3.60 |
| Scenario 3 | | | | | | | | |
| ABOIN | 3.01 | 8.20 | 30.89 | **37.77** | 20.05 | 15.53 | 32.75 | 87.58 |
| BOIN | 0.89 | 7.12 | 29.32 | **40.90** | 21.64 | 23.98 | 45.35 | 0.13 |
| 3+3 | 13.30 | 21.60 | 32.00 | **20.40** | 5.90 | 16.70 | 31.29 | 6.80 |
| Scenario 4 | | | | | | | | |
| ABOIN | 7.47 | 27.71 | **49.83** | 14.39 | 0.44 | 14.49 | 30.75 | 89.31 |
| BOIN | 5.28 | 32.84 | **46.49** | 14.18 | 0.89 | 23.94 | 45.28 | 0.32 |
| 3+3 | 31.20 | 35.50 | **20.70** | 3.10 | 0.10 | 13.70 | 25.51 | 9.40 |
| Scenario 5 | | | | | | | | |
| ABOIN | 21.16 | **44.08** | 27.53 | 5.91 | 0.73 | 13.95 | 29.71 | 89.07 |
| BOIN | 20.57 | **52.57** | 20.74 | 4.39 | 0.65 | 23.81 | 45.04 | 1.08 |
| 3+3 | 41.80 | **29.90** | 7.50 | 0.90 | 0.20 | 11.80 | 21.81 | 19.70 |
| Scenario 6 | | | | | | | | |
| ABOIN | **51.91** | 27.70 | 13.70 | 1.24 | 0.04 | 12.22 | 26.35 | 92.70 |
| BOIN | **55.45** | 25.61 | 5.42 | 0.55 | 0.03 | 22.11 | 41.74 | 12.94 |
| 3+3 | **36.90** | 9.40 | 1.40 | 0.30 | 0.00 | 8.60 | 15.61 | 52.00 |

%MTD: Percentage of MTD selection, Ave SS: Average sample size, Ave SD: Average study duration, %ES: Percentage of early stopping including safety stopping, ABOIN: accelerated BOIN

Table 2 Simulation results for target toxicity level of 25%

| Design | %MTD | | | | | Ave SS | Ave SD | %ES |
|---|---|---|---|---|---|---|---|---|
| | 1 | 2 | 3 | 4 | 5 | | | |
| Scenario 1 | | | | | | | | |
| ABOIN | 0.00 | 0.00 | 0.09 | 0.38 | **99.53** | 14.24 | 30.26 | 99.95 |
| BOIN | 0.00 | 0.00 | 0.00 | 0.01 | **99.99** | 24.00 | 45.40 | 0.00 |
| 3+3 | 0.00 | 0.10 | 0.80 | 1.20 | **97.90** | 18.30 | 34.38 | 0.00 |
| Scenario 2 | | | | | | | | |
| ABOIN | 1.18 | 2.17 | 8.53 | 24.62 | **63.50** | 15.96 | 33.58 | 90.52 |
| BOIN | 0.12 | 0.95 | 4.97 | 26.81 | **67.12** | 23.99 | 45.39 | 0.03 |
| 3+3 | 4.90 | 6.20 | 19.20 | 32.60 | **35.30** | 19.10 | 35.96 | 1.80 |
| Scenario 3 | | | | | | | | |
| ABOIN | 2.27 | 5.70 | 22.02 | **36.60** | 33.36 | 15.70 | 33.09 | 88.26 |
| BOIN | 0.55 | 3.89 | 19.96 | **38.50** | 37.06 | 23.99 | 45.39 | 0.04 |
| 3+3 | 10.10 | 16.00 | 30.20 | **25.70** | 13.10 | 17.60 | 33.03 | 4.90 |
| Scenario 4 | | | | | | | | |
| ABOIN | 4.58 | 19.33 | **47.39** | 25.67 | 2.96 | 14.98 | 31.69 | 88.31 |
| BOIN | 2.40 | 21.30 | **46.13** | 26.27 | 3.77 | 23.98 | 45.35 | 0.13 |
| 3+3 | 24.50 | 31.60 | **28.70** | 7.10 | 0.50 | 14.80 | 27.61 | 7.60 |
| Scenario 5 | | | | | | | | |
| ABOIN | 12.36 | **36.77** | 35.32 | 11.93 | 3.35 | 14.16 | 30.10 | 89.08 |
| BOIN | 10.93 | **42.94** | 30.54 | 11.63 | 3.49 | 23.92 | 45.24 | 0.47 |
| 3+3 | 33.50 | **33.90** | 11.30 | 3.90 | 0.40 | 12.90 | 23.92 | 17.00 |
| Scenario 6 | | | | | | | | |
| ABOIN | **36.06** | 32.13 | 22.92 | 5.44 | 0.47 | 12.87 | 27.62 | 91.61 |
| BOIN | **40.62** | 35.66 | 13.76 | 3.24 | 0.35 | 22.97 | 43.41 | 6.37 |
| 3+3 | **37.90** | 17.40 | 5.20 | 0.90 | 0.00 | 10.20 | 18.72 | 38.60 |

%MTD: Percentage of MTD selection, Ave SS: Average sample size, Ave SD: Average study duration, %ES: Percentage of early stopping including safety stopping, ABOIN: accelerated BOIN

[Accelerated BOIN design]

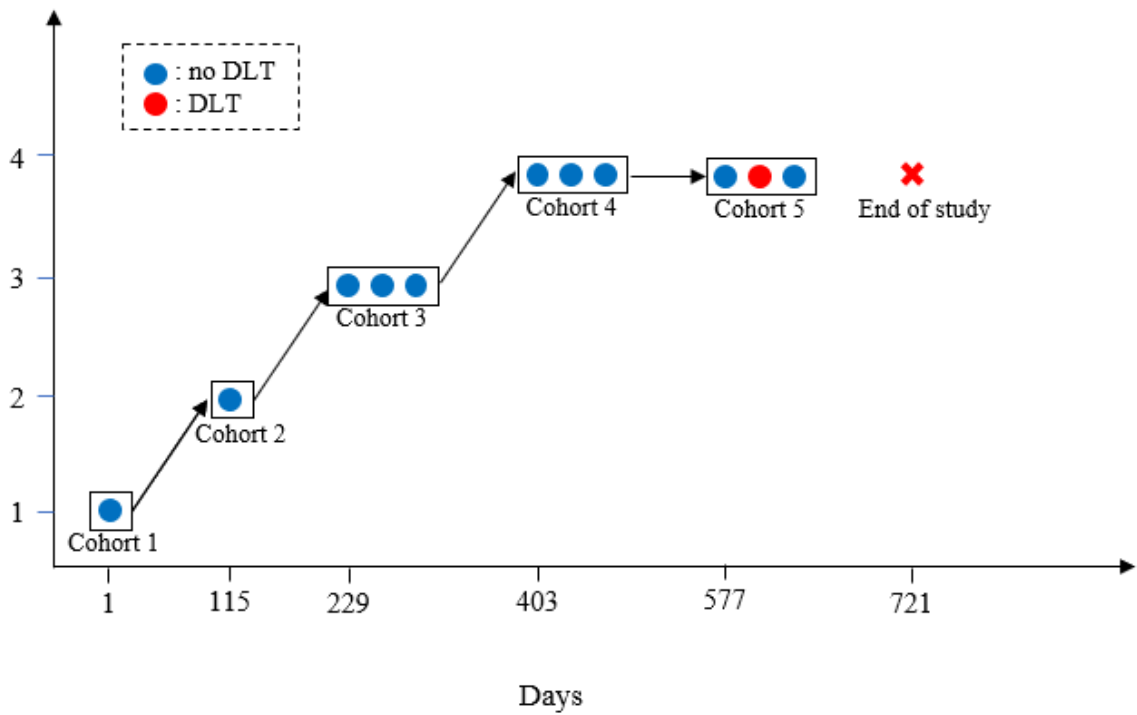

[Standard BOIN design]

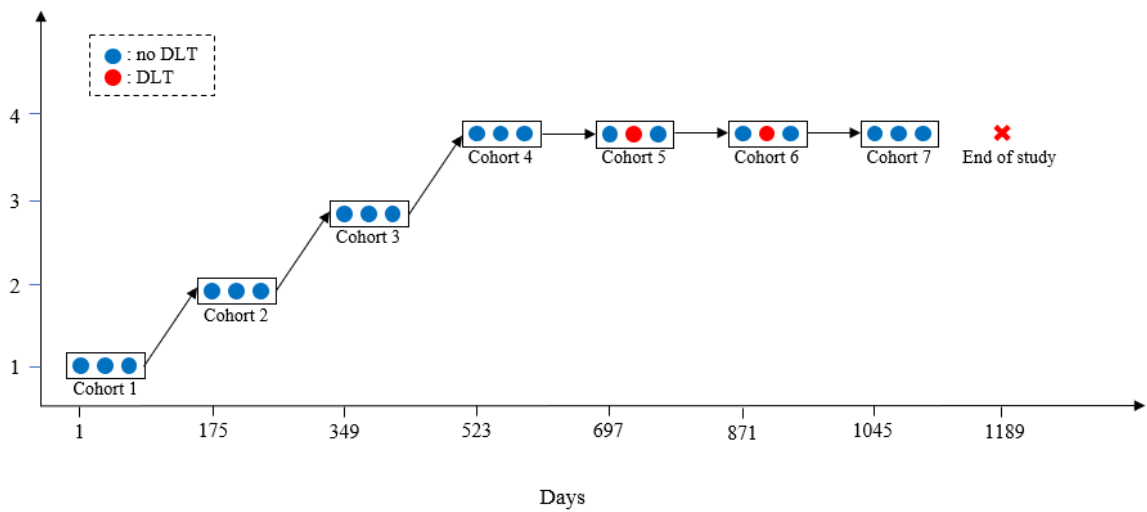

[3+3 design]

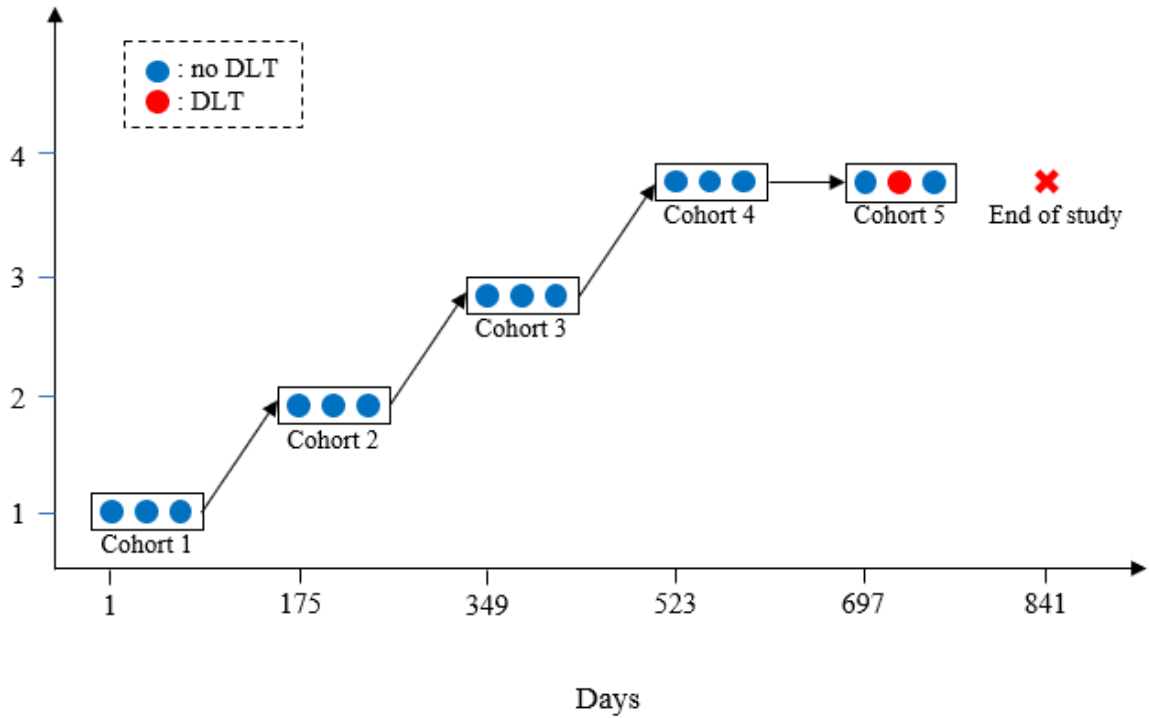

**Figure 1. Dose escalations in accelerated BOIN, standard BOIN, and 3+3 design in case study**

[Percentage of correct MTD selection]

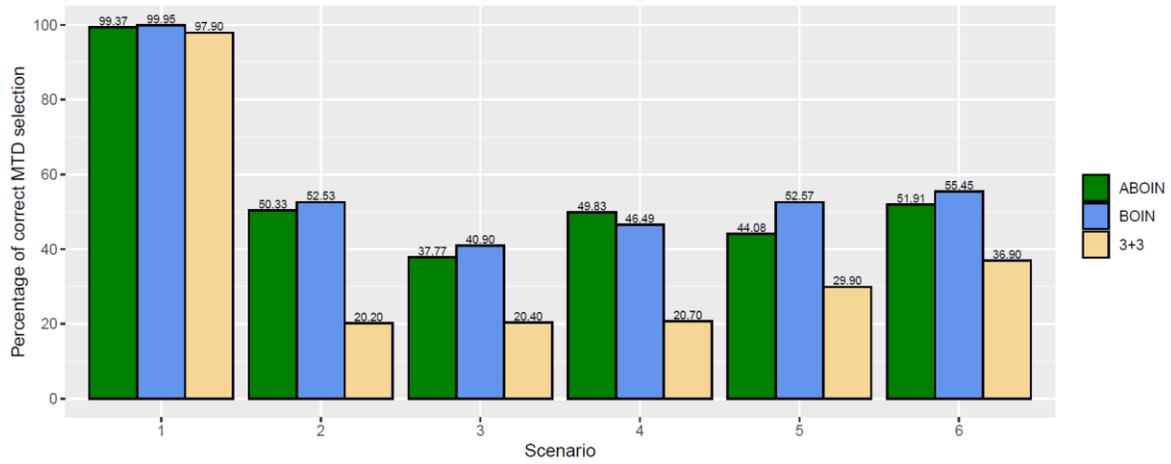

[Average total sample size]

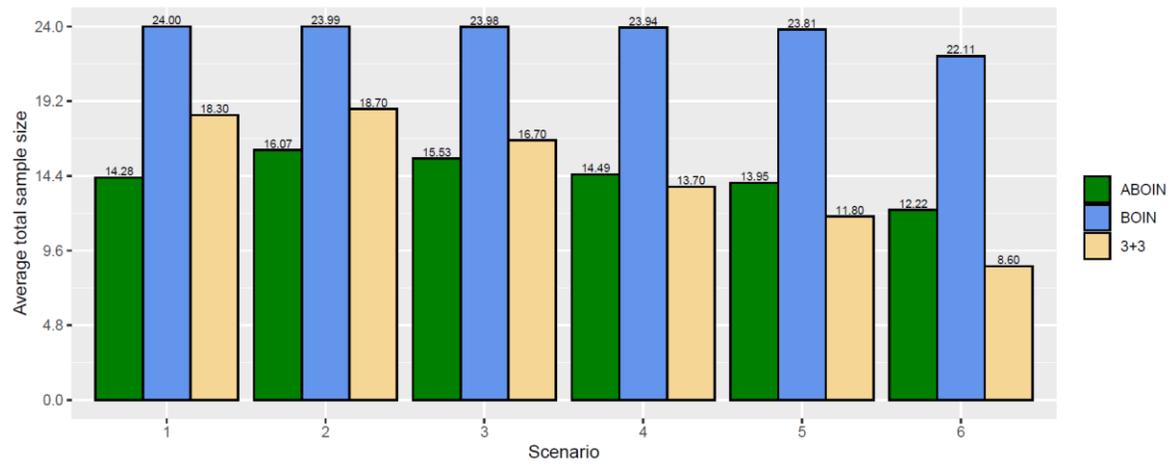

[Average study duration]

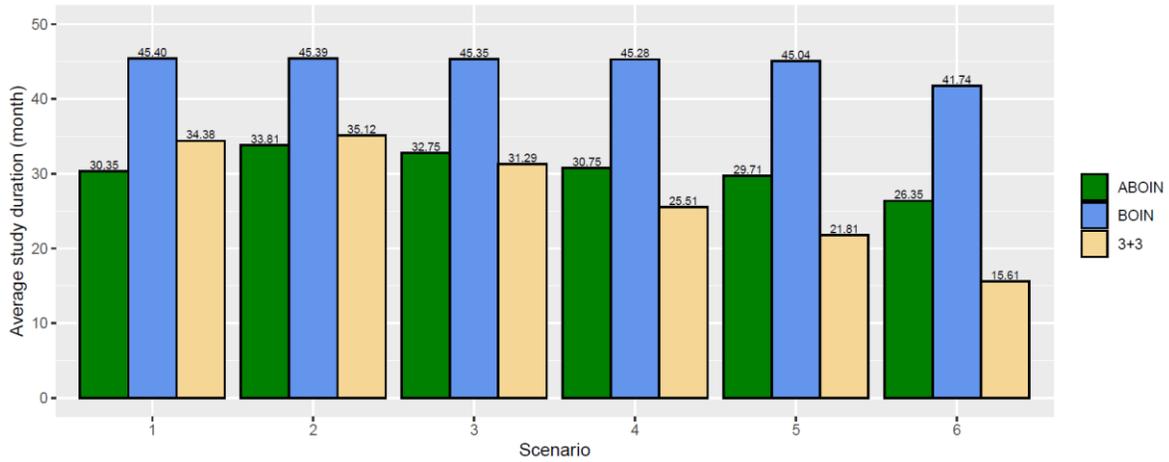

**Figure 2.** Simulation results for target toxicity level of 30%

ABOIN: Accelerated BOIN

[Percentage of correct MTD selection]

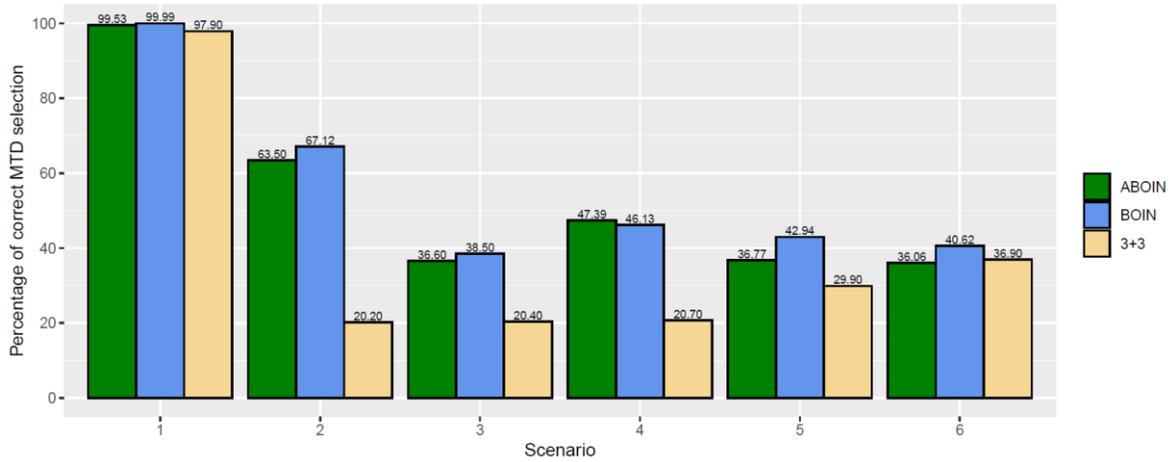

[Average total sample size]

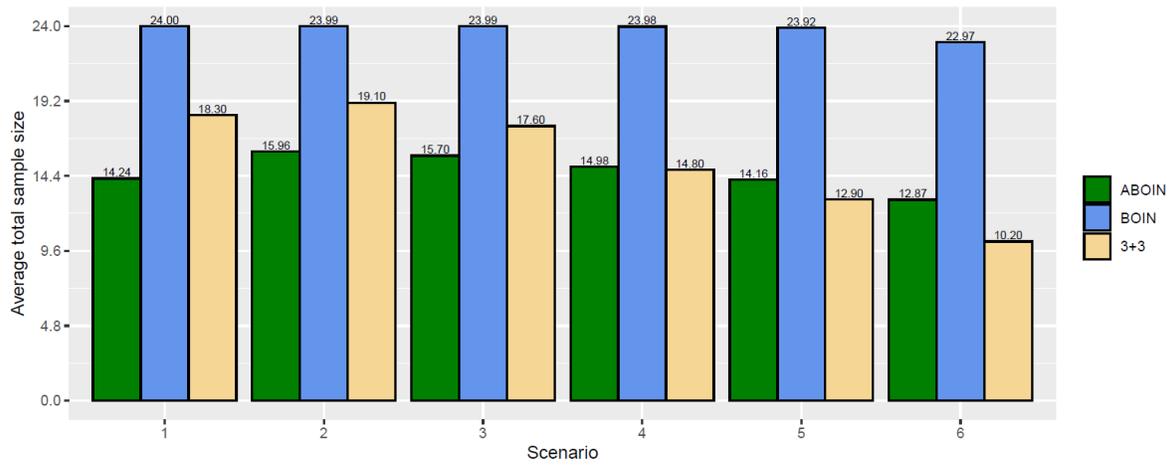

[Average study duration]

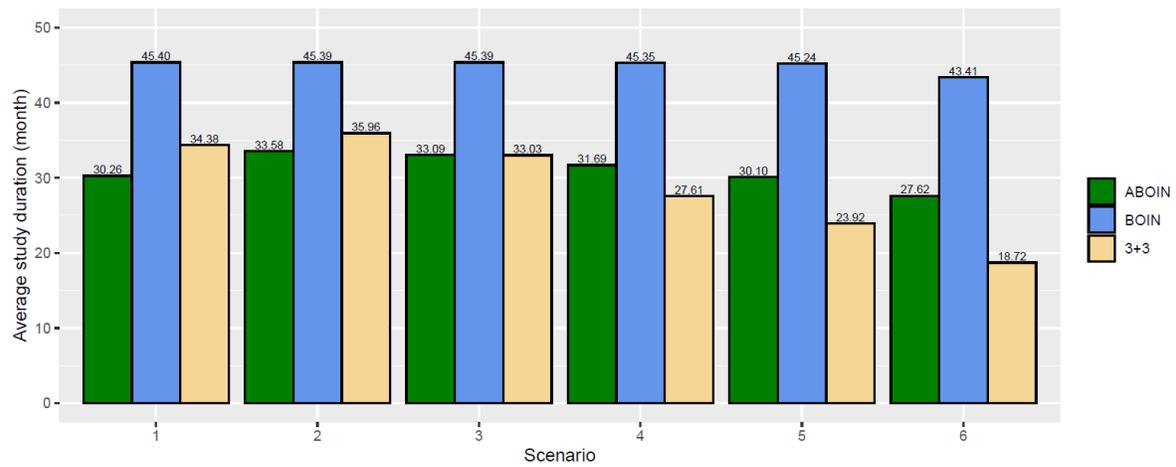

**Figure 3.** Simulation results for target toxicity level of 25%

ABOIN: Accelerated BOIN